\documentclass[journal]{IEEEtran}

\usepackage{amsmath,amssymb,epsfig,epstopdf,siunitx,pgfplots,url,color}

\newcommand{\rev}[1]{#1}
\newcommand{\revre}[1]{#1}

\begin{document}
\bstctlcite{IEEEexample:BSTcontrol}

\title{Adaptive Estimation of the Neural Activation Extent \revre{in Computational Volume Conductor Models of} Deep Brain Stimulation}

\author{Christian~Schmidt\IEEEauthorrefmark{1},
        Ursula~van~Rienen
\thanks{Manuscript submitted to IEEE TBME on April, 26 2017. \newline \textit{Asterisk indicates corresponding author.}}      
\thanks{\IEEEauthorrefmark{1}C. Schmidt is with the Institute
of General Electrical Engineering, University of Rostock, 18059 Rostock, Germany (e-mail: cschmidt18057@gmail.com)}
\thanks{U. van Rienen is with the Institute of General Electrical Engineering, University of Rostock, 18059 Rostock, Germany}}

\maketitle

\begin{abstract}
\textit{Objective:} The aim of this study is to propose an adaptive scheme embedded into an open-source environment for the estimation of the neural activation extent during deep brain stimulation and to investigate the feasibility of approximating the neural activation extent by thresholds of the field solution.
\textit{Methods:} Open-source solutions for solving the field equation in volume conductor models of deep brain stimulation and computing the neural activation are embedded into a Python package to estimate the neural activation \rev{dependent on} the dielectric tissue properties and axon parameters by employing a spatially adaptive scheme. Feasibility of the approximation of the neural activation extent by field thresholds is investigated to further reduce the computational expense.
\textit{Results: } The varying extents of neural activation for different patient-specific dielectric properties were estimated with the adaptive scheme. The results revealed the strong influence of the dielectric properties of the encapsulation layer in the acute and chronic phase after surgery. \rev{The computational time required to determine the neural activation extent in each studied model case was substantially reduced.} 
\textit{Conclusion: } The neural activation extent is altered by patient-specific parameters. Threshold values of the electric potential and electric field norm facilitate a computationally efficient method to estimate the neural activation extent.
\textit{Significance: } The presented adaptive scheme is able to robustly determine neural activation extents and field threshold estimates for varying dielectric tissue properties and axon diameters while reducing substantially the computational expense.
\end{abstract}

\begin{IEEEkeywords}
Deep brain stimulation (DBS), Finite element methods, Neural activation, Open-source
\end{IEEEkeywords}

\IEEEpeerreviewmaketitle
\IEEEPARstart{D}{eep brain stimulation} (DBS) is a widely employed effective procedure to treat symptoms of motor disorders such as Parkinson's disease (PD), essential tremor and dystonia \cite{benabid2009, johnson2013} and consists in the implantation of an electrode lead into deep brain target areas. A common target in DBS is the subthalamic nucleus (STN), which constitutes a preferred target for the treatment of PD. The STN consists of different functional zones, which are classified into limbic, associative, and \rev{sensorimotor} zones \cite{guigoni2005}, from which electrical stimulation of the \rev{sensorimotor} zone is mostly associated with the relief of motor symptoms of PD \cite{follett2010}. Due to patient-specific parameters, such as brain structure anatomy, dielectric tissue properties, electrode location, and severity of symptoms, the adjustment and optimization of stimulation parameters during and after surgery can be rather time consuming and connected to additional costs. Computational models provide a possibility to estimate the stimulation impact by determining activated areas in the deep brain based on the given patient-specific parameters. The extent of neural activation, or the volume of tissue activated (VTA), is a common computational modeling approach to estimate the size of the activated tissue during DBS and has been applied in various computational studies in this area including homogeneous \cite{butson2005}, rotationally symmetric \cite{butson2006}, heterogeneous \cite{grant2010, schmidt2016}, and anisotropic volume conductor models \cite{schmidt2012} of the human brain and deep brain target areas. In general, the approach is based on positioning a number of \rev{models of mammalian nerve fibers (axon models)} in a grid located in a plane perpendicular to the electrode lead. For each axon model the computational goal of finding the minimum stimulation amplitude required to activate the axon is solved. From the resulting threshold values at the grid points, a threshold isoline for a given stimulation amplitude is determined. This procedure is repeated for multiple planes rotated around the electrode lead. In case of a rotationally symmetric field model, it is sufficient to compute the threshold isoline in one plane and revolve the solution around the electrode lead. The resulting threshold isolines then provide the measure for computing the VTA. The drawback of the method with respect to computational ressources and adaptivity is that the location of the axon nodes to include a range of desired stimulation amplitudes has to be available prior to the simulation, which involves several pre-simulation runs, which often are carried out manually. The field solution in the target area is commonly computed by creating a volume conductor model of the DBS electrode and the surrounding tissue. For solving the governing equations, which are typically the stationary current field or electro-quasistatic equation for DBS applications \cite{grant2010, bossetti2008}, often commercial software solutions, such as COMSOL Multiphysics\textsuperscript{\textregistered} (http://www.comsol.com) are used \cite{butson2005, grant2010, schmidt2016, schmidt2013, mcintyre2004, astrom2015}, \rev{while, to our knowledge, no studies employing open-source solutions on the field model have been published in scientific journals yet.} Regarding the coupling of the neuronal activation in axon models and the extracellular field distribution, a Python package with the purpose to compute the local field potentials for a given axon distribution of defined activation was presented in \cite{linden2013}. To date, no open-source solution to model the field distribution during DBS and to estimate the resulting neural activation extent exists.\newline \indent
The computation of the VTA is a computationally demanding task. In a previous computational study investigating the relationship between the neural activation and the field solution, an approximation of the extent of neural activation by threshold values of the field solution and their derivates have been investigated \cite{astrom2015}. The results suggested that electric field norm thresholds are a good estimator for the extent of neural activation. In such a case only a small neural activation extent has to be computed to get the initial field norm threshold estimate, from which the neural activation extents for varying stimulation amplitudes can be derived. Especially for large diameter axon fibers, the electric field norm constituted a good estimate. The relationship between electric field norm and neural activation was determined by positioning axons normal to the electrode lead along a line originating at the active electrode contact center in a homogeneous volume conductor model for DBS. The threshold value of the electric field norm was equivalent to its value at the maximum neural activation distance for a given stimulation amplitude. The assumption of this approach is that the shape of the neural activation extent is spherical, which is true for a homogeneous volume conductor model and a spherical or point source. For heterogeneous volume conductor models and DBS electrode geometries, the shape deviates from the spherical shape.\newline \indent
The goal of this study is to investigate the feasibility of approximating the neural activation extent by the field solution of a heterogeneous volume conductor model for DBS incorporating a DBS electrode, encapsulation layer, brain tissue, and the STN as target area. To automate the task of determining the neural activation extent and further reduce the computational demand, an algorithm is proposed which determines adaptively the location of axons being activated within a defined stimulation amplitude range or distance range. Besides dropping the need for manually determining the number of axons in the target area for a given stimulation amplitude range, the approach reduces the computational expense by omitting the computation of activation for axons which are located outside the activation volume. The model pipeline for the computation of the field solution and the neural activation is implemented in a Python package and embedding open-source tools for the model generation, meshing, and solving. The field solution and the neural activation are validated using analytical models as well as reference data published in literature. The Python package is designed modular, which allows to interchange field as well as neuron models and to adjust model parameters accordingly. The Python package, as well as the code to replicate the data and figures of this study are made available open-source\footnote{\rev{\url{https://bitbucket.org/ChrSchmidt83/fanpy/get/fanpy-1.2.zip}}}.
\section{Methods}

\subsection{Model geometry}
\begin{figure}[t]
    \centerline{\psfig{figure=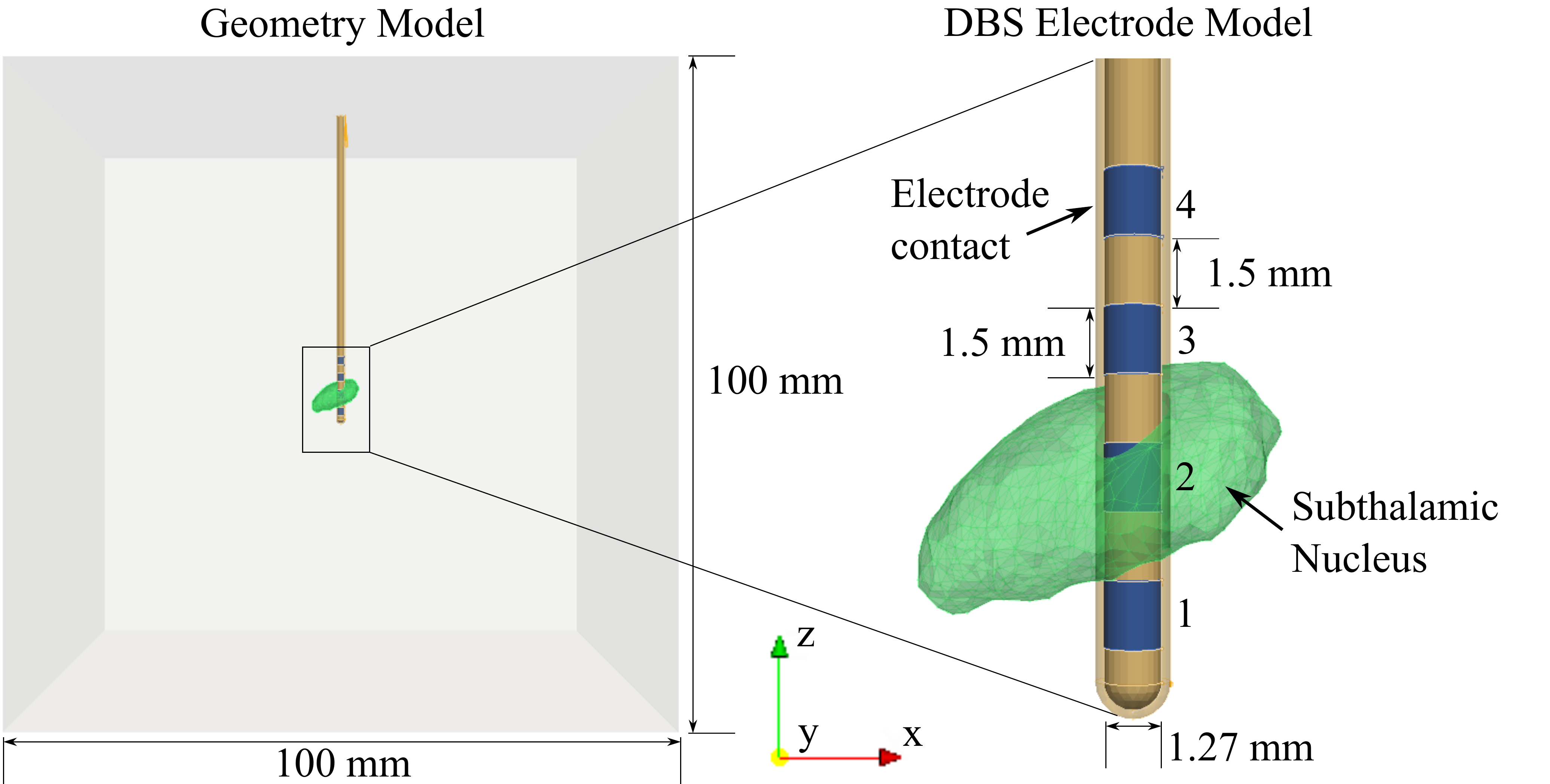,width=1.0\columnwidth} }
    \caption{The model geometry consists of a DBS electrode, an encapsulation layer around the DBS electrode, and a STN model. The model compartments are surrounded by a bounding box.}
    \label{fig:modelgeometry}
\end{figure}

Following the approach in \cite{mcintyre2004a}, the model geometry consists of a DBS electrode model located in a bounding box comprising the different tissue compartments. The geometry of the DBS electrode represents a Medtronic lead model 3387 (Medtronic, Inc., Minneapolis, MN). To account for the inflammatory response of the body tissue to the electrode implant, an encapsulation layer with a thickness of \SI{0.2}{mm} was incorporated around the electrode body. The bounding box size was determined by an edge length of \SI{100}{mm}. The geometry model of the STN based on a functional zones atlas \cite{accolla2014} was generated by \rev{creating} a surface model out of the right STN threshold maps of the atlas using the open-source software platform 3D Slicer (\url{https://www.slicer.org/}, version 4.2.1). The whole geometry model was generated with the open-source software SALOME (\url{http://www.salome-platform.org/}, version 7.8.0). After creating and merging of the different compartments, the STN surface model was converted to a solid and positioned in the geometry with the second electrode contact of the DBS electrode located in the \rev{sensorimotor} zone (Fig. \ref{fig:modelgeometry}).

\subsection{Manual mesh refinement and subdomain generation}
\begin{figure}[t]
    \centerline{\psfig{figure=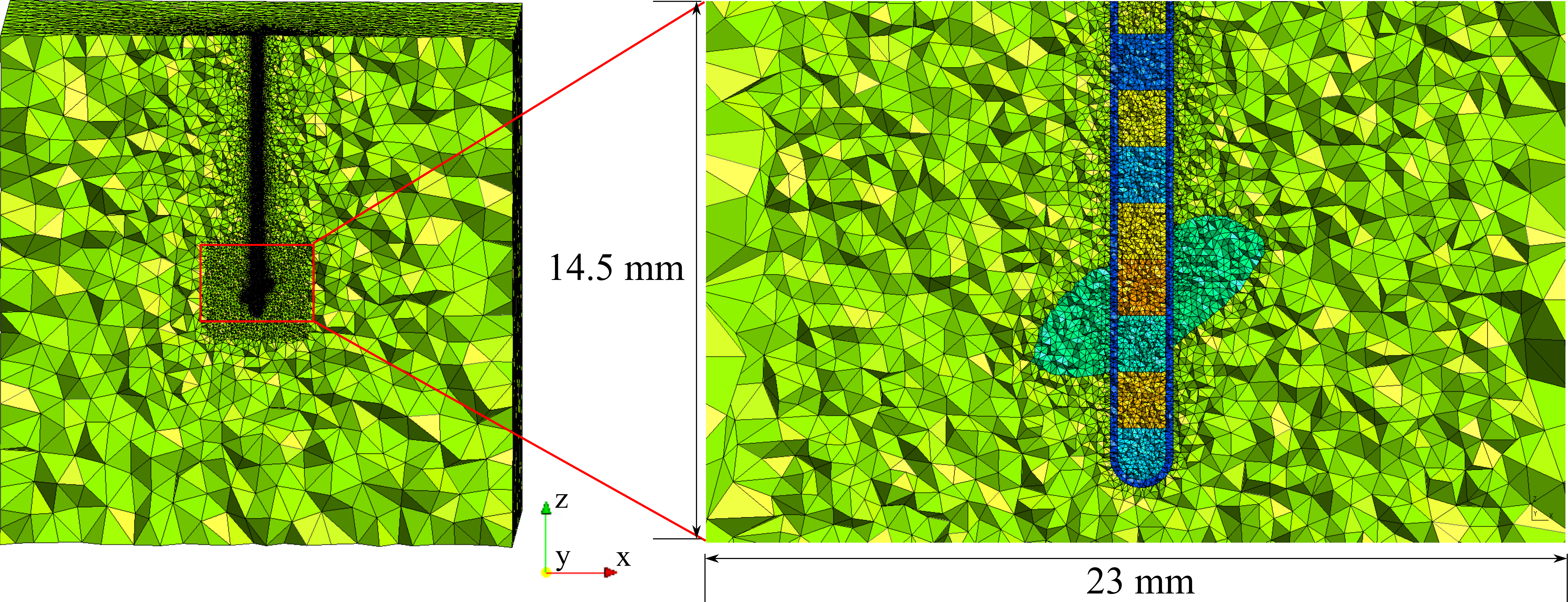,width=1.0\columnwidth} }
    \caption{Manually refined mesh of the computational domain in the $xz$-plane for $y=0$. The different computational subdomains including the bounding box, electrode lead, electrode contacts, encapsulation layer, and STN, are shown with different colors.}
    \label{fig:mesh}
\end{figure} 
The meshing of the computational domain is carried out with the open-source mesh generator Gmsh (\url{http://gmsh.info}, version 2.10.1). Therefore, the model geometry is exported from SALOME in the \texttt{brep} format and loaded into gmsh. Since the geometry contains entities of varying scales (Fig. \ref{fig:modelgeometry}), the mesh was manually refined at the surfaces of the entities by specifying a characteristic length of the finite elements. Based on values for the manual mesh refinement for DBS volume conductor models of the human brain \cite{schmidt2012}, which provided a sufficient refinement of the computational domain, the characteristic length was set to \SI{0.1}{mm} for the electrode lead, electrode contacts, and encapsulation layer, to \SI{0.2}{mm} for the STN, and \SI{5.0}{mm} for the bounding box. Additionally, a refined cubical mesh region with an edge length of \SI{20}{mm} and a characteristic length of \SI{0.5}{mm} was defined in the target area. The final mesh contained approximately $240{,}000$ vertices and $1.5$ million cells (Fig. \ref{fig:mesh}). In order to assign material properties and boundary conditions to the model compartments and surfaces, the subdomains of the geometry model were assigned and grouped to physical volumes and surfaces. The information on the manual mesh refinement and the defined physical volumes and surfaces is stored in a \texttt{geo} file, from which automatically the mesh can be generated.

\subsection{Incorporating the model into FEniCS}
The mesh generated with Gmsh is converted into a format readable by the open-source simulation software \mbox{FEniCS} (\url{https://fenicsproject.org}, version 2016.2.0) by using the command line tool \texttt{dolfin-convert}. To faciliate the interchange of different model geometries and meshes, information on the model's subdomains, boundaries, and default boundary conditions and material properties are defined in an \texttt{xml} file. The FieldModel module of the designed Python package loads the definitions in the model \texttt{xml} file, checks the information for consistency, generates the mesh and applies the material properties and boundary conditions. While the conductivity of the electrode lead (insulation) and the electrode contacts (platinum-iridium) are kept constant with a value of \SI[exponent-product = \cdot]{1e-7}{Sm^{-1}} for the electrode lead, and \SI[exponent-product = \cdot]{1e7}{Sm^{-1}} for the electrode contacts, the conductivity values for the encapsulation layer, brain tissue, and STN varied for the different study cases. The field equation is determined by a stationary current field problem, which is commonly applied in various computational modeling studies for DBS \cite{astrom2015, mcintyre2004a}:
\begin{equation}
	\nabla \cdot \left[\sigma (\boldsymbol r) \nabla(\varphi (\boldsymbol r))\right] = 0 , \boldsymbol r \in \Omega
	\label{eq:laplace}
\end{equation}
with the conductivity $\sigma$, the electric potential $\varphi$ and the computational domain $\Omega$. If the capacitive and dispersive properties of human tissue are taken into account, the field equation (\ref{eq:laplace}) has to be reformulated for complex materials with
\begin{equation}
	\sigma_c (\omega, \boldsymbol r) := \sigma(\omega, \boldsymbol r) + \mathsf{j}\omega\epsilon_0\epsilon_r(\omega, \boldsymbol r)
	\label{eq:sigmac}
\end{equation}
with the imaginary unit $\mathsf{j}$, the angular frequency $\omega$, the electric field constant $\epsilon_0$, and the relative electric permittivity $\epsilon_\mathsf{r}$, which placed in equation (\ref{eq:laplace}) resembles a quasistatic field problem. The field equations are solved with the finite element method by formulating the variational problem within FEniCS. Dirichlet boundary conditions are applied to the surface of the second electrode contact, located within the STN (Fig. \ref{fig:modelgeometry}), with a unit value of \SI{1.0}{V}, and the exterior boundary of the bounding box with a value of \SI{0.0}{V}, serving as ground. The resulting linear system of equations was solved for quadratic nodal basis functions using the generalized minimal residual method with a relative tolerance of $1\cdot 10^{-6}$ and an absolute tolerance of $1\cdot 10^{-7}$, employing an algebraic multigrid preconditioner in case of the stationary current field problem. It was ensured that the deviation in the electric potential and electric field norm in the prescribed activation distances (see section \ref{sec:approxvta}) as well as the impedance of the model was below \SI{1}{\percent} if cubic ansatz functions were employed. Plots of the field distribution were visualized with the open-source visualization application Paraview (\url{http://www.paraview.org/}, version=5.0.1).

\subsection{Electrical properties of human tissue}
\label{sec:tissueproperties}
The conductivity and relative permittivity of biological tissue show a frequency dependence which can be described by different dispersion regions and parametrized by assembled Cole-Cole equations \cite{gabriel1996} representing a complex conductivity
\begin{equation}
\sigma_c(\omega) =  \epsilon_\infty + \frac{\sigma_\mathsf{ion}}{\mathsf{j} \omega  \epsilon_0} + \sum_{i=1}^4 \frac{\Delta \epsilon_i}{1+(\mathsf{j} \omega \tau_i)^{1-\alpha_i}}
\label{eq:colecole}
\end{equation}
with the static ionic conductivity $\sigma_\mathsf{ion}$, the relaxation time constants $\tau_i$, the dispersion constant $\alpha_i \in [0,1]$, and the relative permittivity at high frequency $\epsilon_\infty$  as well as the difference of the low and high frequency relative permittivity $\Delta \epsilon_i$. The tissue model parameters were taken for white matter, grey matter, and cerebrospinal fluid from \cite{gabriel1996}. The electrical tissue properties of the encapsulation layer vary over time from an acute phase immediately after surgery to a chronic phase after some weeks due to cell growth in the layer \cite{grill1994}. The acute phase was modeled by the dielectric properties of cerebrospinal fluid \cite{grant2010}, while the chronic phase was modeled by dividing the values for white matter by a factor of 2 \cite{mcintyre2004a}.

\subsection{Voltage-controlled and current-controlled stimulation}
The time-dependent electrical potential in the target area for a given stimulation signal was determined by forming the outer product of the field solution for a unit voltage set to the active second electrode contact (Fig. \ref{fig:modelgeometry}) and the voltage- or current-controlled stimulation signal. The stimulation signals commonly applied in DBS therapy in humans consist of a monophasic square-wave signal with pulse durations in the range of \SI{60}{\micro s} - \SI{100}{\micro s} and a repetition frequency in the range of \SI{130}{Hz} - \SI{150}{Hz} \cite{mcintyre2004a, astrom2015, grant2010, gimsa2005}. To avoid charge accumulation in the tissue, the monophasic stimulation pulse is often followed by a reversed charge-balancing pulse of substantially smaller amplitude compared to the active stimulation pulse. Considering that the activation of a neuron is mainly influenced by the amplitude of the stimulation pulse \cite{grant2010}, the reversed charge-balancing pulse is not considered in this study. While the time-dependent electrical potential in the target area for voltage-controlled stimulation is provided by the outer product of the field solution for a unit voltage and the voltage-controlled stimulation signal, the time-dependent electrical potential for current-controlled stimulation requires an additional scaling of the field solution by the electrode impedance, which is computed by dividing the square of the unit voltage by the electric power $P$ of the field model.
\begin{equation}
	P = \int \limits_\Omega \langle \sigma(\boldsymbol r) \nabla \varphi(\boldsymbol r), \nabla \varphi(\boldsymbol r) \rangle \, \mathsf{d}x
\end{equation}
with the inner product $\langle,\rangle$, corresponding to a scaling of the unit voltage at the active electrode contact boundary condition by a factor equal to a unit current flowing through its surface.

\subsection{Neuronal activation model}
The neural activation model is based on a myelinated axon cable model, which includes $21$ nodes of Ranvier, paranodal and internodal segmenets as well as the myelin sheath \cite{mcintyre2002}, following the assupmtion that activation occurs along the axon \cite{mcintyre2004}. Based on the model parameters given in \cite{mcintyre2002}, the model is parametrized with respect to the fiber diameter comprising nine distinct diameters between \SI{5.7}{\micro m} and \SI{16.0}{\micro m}\rev{, which were extented by the parameters for \SI{2.0}{\micro m} and \SI{3.0}{\micro m} fiber diameter taken from \cite{sotiropoulos2007}}. The model is implemented in the open-source simulation environment NEURON (\url{https://www.neuron.yale.edu/neuron/}, version=7.4)\footnote{\url{https://senselab.med.yale.edu/modeldb/showModel.cshtml?model=3810}} with the time-dependent electrical potential for an applied stimulation signal at the location of each axon compartment applied to its extracellular mechanism node neglecting any axon contribution to the extracellular field distribution \cite{mcintyre2004a}. The axon activation is determined by solving the linear system of differential equations resulting from the membrane dynamics with the backward Euler method for a time step of \SI{5}{\micro s} \cite{mcintyre2004a}. An axon was considered to be activated when the inner potential at the exterior nodes of Ranvier of the model obtained a threshold value of larger than \SI{0}{mV}, representing a generated spike as result of the stimulation pulse, in a $1$-to-$1$ ratio for $10$ delivered stimulation pulses.

\subsection{Adaptive estimation of the neural activation extent}
\begin{figure}[t]
    \centerline{\psfig{figure=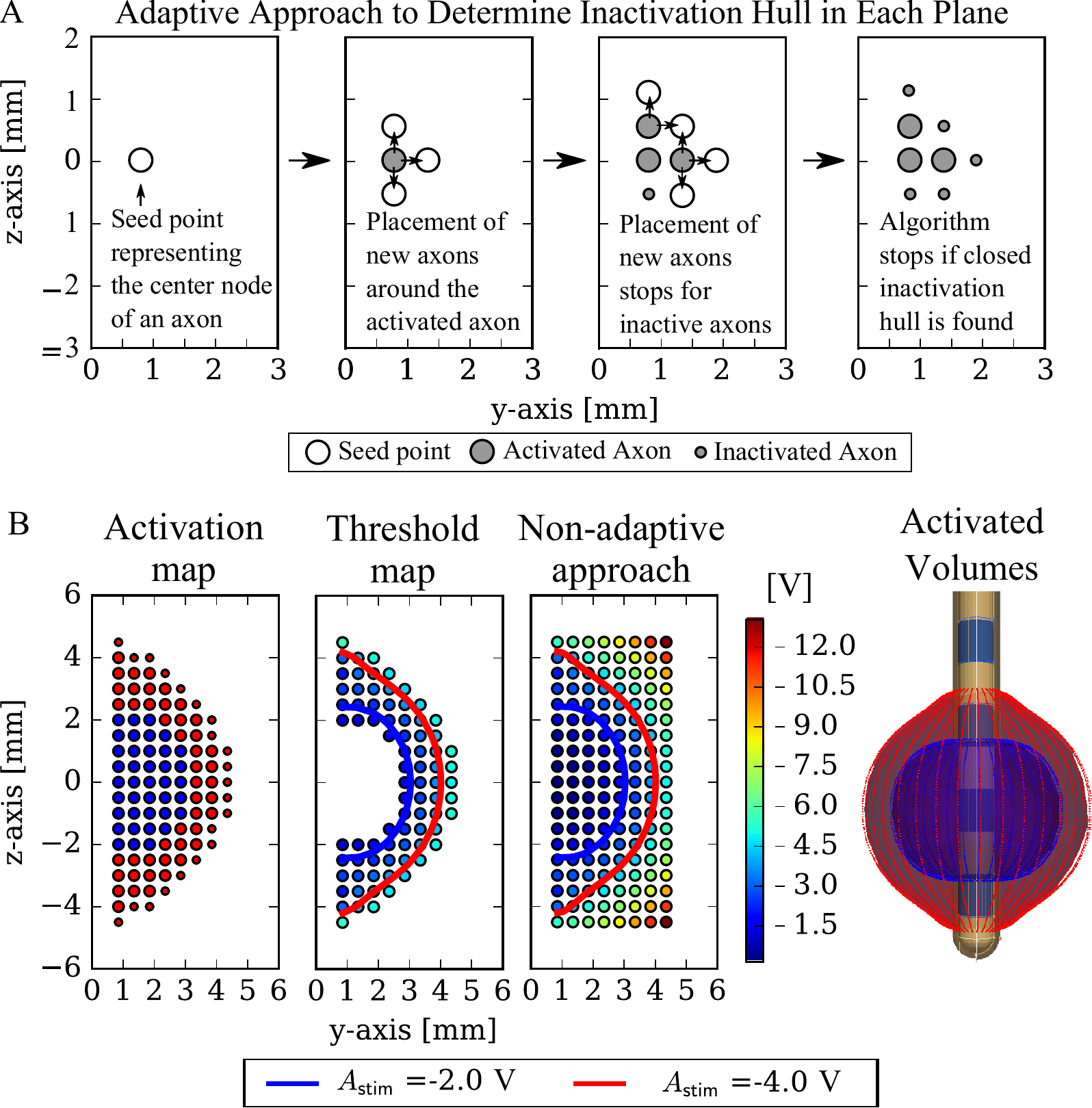,width=1.0\columnwidth} }
    \caption{\rev{Illustration of the adaptive algorithm for the estimation of the neural activation extent in the range of given stimulation amplitudes. A: Seed points are placed in front of the active stimulation electrode (here one seed point for monopolar stimulation). If the axon is activated by the stimulation new seed points are placed around the activated axon. If the axon is not activated, the algorithm stops to place new seed points around this axon. The procedure is continued until a closed hull of inactivated axons is found. B: For a given stimulation amplitude range, the algorithm determines closed activation hulls for the minimum and maximum stimulation amplitude and removes unneeded interior points. For a given tolerance, the \rev{minimally} required stimulation amplitude is computed at each point resulting in a threshold map. From this map, the activation isolines can be computed for any stimulation amplitude in the given range. The procedure is repeated for several planes around the stimulation electrode lead, from which the extent of neural activation is finally estimated. \revre{The threshold map computed for the non-adaptive approach is shown as comparison.}}}
    \label{fig:vta_adaptive}
\end{figure}

The proposed algorithm determines adaptively the required location of axons in the target area for a given range of stimulation amplitudes, which ensures that the extent of neural activation for any stimulation amplitude within the given range can be computed from the determined axon locations. \rev{The required axon locations are determined in cutting planes, which are located around the electrode lead. First, a seed point, located \rev{at} a distance of \SI{0.85}{mm} to the active electrode contact's center, which corresponds to the extent of the electrode with the encapsulation thickness, is placed in such a plane. Next, the algorithm determines whether the axon\rev{, which is positioned perpendicular and centered to the electrode lead at the seed point location,} is activated for the minimal given stimulation amplitude. If an activation was recorded, the algorithm continues by placing further axons around the activated axon, with their center node location positioned radial ($\varrho+\Delta s$) and parallel ($z+\Delta s, z-\Delta s$) to the electrode lead with a \rev{step size} $\Delta s$, and determining their activation.} If the axon was not activated, the algorithm stops for this location. The procedure is continued until an inactivated hull of axons is determined\rev{, which is achieved if in each line \rev{radial} to the electrode the axon furthest away to the electrode is not activated by the stimulation}. This inactivated hull is then used as seed points for determining the axon locations for the maximum stimulation amplitude within the given range by applying the same algorithm. Finally, \rev{the} interior points, which are located inside the activated hull of axons for the minimum given stimulation amplitude are removed from the set of locations, resulting in a shell of axon locations for the given stimulation amplitude range. \rev{Determining the minimally required stimulation amplitude to elicit an action potential in the axons at these locations represents a root-finding problem, which is solved using the bisection method (binary search method) with a tolerance of $1\cdot 10^{-6}$} (Fig. \ref{fig:vta_adaptive}). The algorithm is carried out for \rev{$n_\alpha = \lceil360/\Delta \alpha\rceil$} planes around the electrode with the rotational degree \rev{step size} \rev{$\Delta \alpha$}. The subsequent computation of the neural activation for each axon model introduced by the algorithm as well as finding the \rev{minimally} required stimulation amplitude to activate the axon model are carried out by the model pipeline in parallel \rev{with worker threads adding and withdrawing axon points to a pool (Python Queue) of axons. In order to prevent the placement of new axon points at the same location to the pool, a new axon point is only added if an axon point with the same location was not added before by another worker thread.}

\subsection{Approximating the neural activation by field thresholds}
\label{sec:approxvta}
\begin{table}[!t]
	\centering
  \renewcommand{\arraystretch}{1.3}
	\caption{\label{tab:models}Conductivity of the volume conductor compartments for the different study cases. The values were determined by using the tissue parameters from \cite{gabriel1996}.}
	\begin{tabular}{lccccc}
			\hline
			\bfseries Study Case & \multicolumn{3}{c}{\bfseries Conductivity  [\SI{}{Sm^{-1}}]} \\
			& \bfseries encapsulation & \bfseries brain & \bfseries subthalamic \\
			& \bfseries layer & \bfseries tissue & \bfseries nucleus (STN) \\
			Model 1 (homogeneous) & 0.064 & 0.064 & 0.064 \\
			Model 2	(acute phase) & 2.000 & 0.064 & 0.064 \\
			Model 3	(chronic phase) & 0.032 &	0.064 & 0.064 \\
			Model 4 (with STN) & 0.032 & 0.064 & 0.103 \\
			\hline
	\end{tabular}	
\end{table}
The proposed adaptive algorithm is used to assess the feasibility of approximating the neural activation extent for various stimulation amplitudes by threshold values of the electrical potential and the electric field norm. A current-controlled stimulation is applied, which is gaining increased interest in clinical application due to the reduced side effects and sensitivity to inter-individual \rev{variabilities} \cite{lempka2010, lettieri2015}. Different values for the dielectric properties of the volume conductor model compartments are employed to investigate the approximation quality for different cases of tissue heterogeneity in the target area. The dielectric properties are obtained from equation (\ref{eq:colecole}) using the parameters for white matter, grey matter, and cerebrospinal fluid from \cite{gabriel1996} at a frequency of \SI{2}{kHz}, which constitutes a good approximation of the dispersive nature of the tissue properties for common DBS signals \cite{schmidt2016}. The corresponding conductivity values are approximately \SI{0.064}{Sm^{-1}} for white matter, \SI{0.103}{Sm^{-1}} for grey matter, and \SI{2.000}{Sm^{-1}} for cerebrospinal fluid. The cases comprise a homogeneous model (Model 1) with the conductivity of the encapsulation layer, the brain tissue, and STN set to the value for white matter, a model with a high conductive encapsulation layer (Model 2) set to the value of cerebrospinal fluid \cite{grant2010} representing the acute phase, as well as with a \rev{low} conductive encapsulation layer (Model 3), set to half the value of white matter \cite{mcintyre2004} representing the chronic phase, and a heterogeneous model (Model 4) with a low conductive encapsulation layer, brain tissue set to the value of white matter, and the STN set to the value of grey matter. Since the varying dielectric tissue properties result in a variation of the conduction in the volume conductor model, the extent of neural activation is dependent on the tissue properties \cite{schmidt2013}. \rev{Therefore, DBS with the same stimulation amplitude results in different sizes of the VTA depending on the tissue properties in the models. In order to ensure a stimulation of the target region between a distance of \SI{2.0}{mm} (activation of \rev{sensorimotor} functional zone) and \SI{4.0}{mm} (activation of larger parts of the STN) from the electrode center, the required stimulation amplitude to activate a homogeneous volume within this minimum and maximum distance was computed for each model in advance.} The determinted stimulation amplitude range is then prescribed to the adaptive algorithm to constrain the extent of the estimated neural activation. Except for the latter model, the neural activation extent is computed exploiting rotational symmetry by computing the extent in a reference plane and revolving the solution around the electrode. A spatial \rev{step size} of \rev{$\Delta s=0.5\,\mathsf{mm}$} and a rotational \rev{step size} of \rev{$\Delta \alpha=10\,$\si{\degree}}. \rev{The volumes of the neural activation extent and the corresponding extents determined by the threshold values of the electric potential and the electric field norm were computed by using the Qhull library (\url{http://www.qhull.org/}) implemented in SciPy (\url{https://www.scipy.org/}, version 0.17.0).\newline \indent
A previous computational study investigated the feasibility of approximating the neural activation extent determined by the coupling of the field distribution with axon models by using constant field threshold values, depending on the stimulation protocol and the axon diameter \cite{astrom2015}. The used axon models were based on a multi-compartment mathematical model employing the cable equation, while in this study a double cable axon model is applied \cite{mcintyre2002}. The results showed that iso-volumes for threshold values determined using the electric field norm \rev{allowed for a close approximation of} the neural activation extent determined from the coupled field-axon models. The deviation between these iso-volumes and the neural activation extents further decreased with increasing fiber diameter of the axons. To determine these threshold values, axon models were used to compute the maximum distance along a line \rev{radial} to the active electrode center, for which an axon model placed at this distance is still activated by the stimulation. The electric potential or the electric field norm at this distance provided the field threshold value for computing the neural activation extent. This procedure was repeated for a set of stimulation amplitudes in a given range, resulting in averaged field threshold values over the given stimulation amplitude range. The neural activation volumes resulting from this approach and using the electric field norm to determine the field threshold values are denoted as VTA$_{E,\mathsf{const}}$ in this study. For evaluating the feasibility of this approach in \cite{astrom2015}, the threshold values for a given stimulation amplitude range were normalized with respect to the threshold value for the minimum stimulation amplitude.\newline \indent
In addition to this methodology, this study incorporates an approach, which does not average the field threshold values in the given stimulation amplitude range, but computes the threshold-distance relationship along the line \rev{radial} to the active electrode center, from which the maximum distance to activate an axon for a given stimulation amplitude can be computed in post-processing. In addition, the proposed methodology accounts also for rotationally asymmetric field distributions by computing the threshold-distance relationship in each plane around the stimulation electrode lead and averaging the resulting field threshold values. The corresponding iso-volumes are denoted as VTA$_\varphi$ if electric potential threshold values and VTA$_E$ if electric field norm threshold values were applied.} To investigate the dependence of the resulting approximation on the fiber diameter as reported in \cite{astrom2015}, the proposed study cases are carried out for axon fibers with diameters of \rev{\SI{2.0}{\micro m}, \SI{3.0}{\micro m},} \SI{5.7}{\micro m}, \SI{7.3}{\micro m}, \SI{8.7}{\micro m}, an \SI{10.0}{\micro m}. Following the approach in \cite{astrom2015}, the thresholds are determined by the mean value of the electric potential and the electric field norm at the activation distance \rev{radial} to the center of the active electrode contact for a given stimulation amplitude within the prescribed range in each plane. 

\section{Results}
\subsection{Validation of the field solution}
\begin{figure}[t]
    \centerline{\psfig{figure=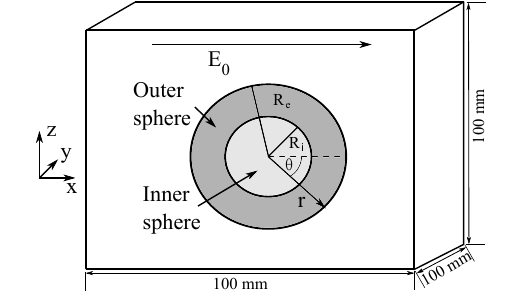,width=0.6\columnwidth} }
    \caption{Geometry of the analytical validation example consisting out of a layered sphere located in an homogeneous electric field $E_0$.}
    \label{fig:analyticalproblemgeometry}
\end{figure}
\begin{figure}[t]
    \centerline{\psfig{figure=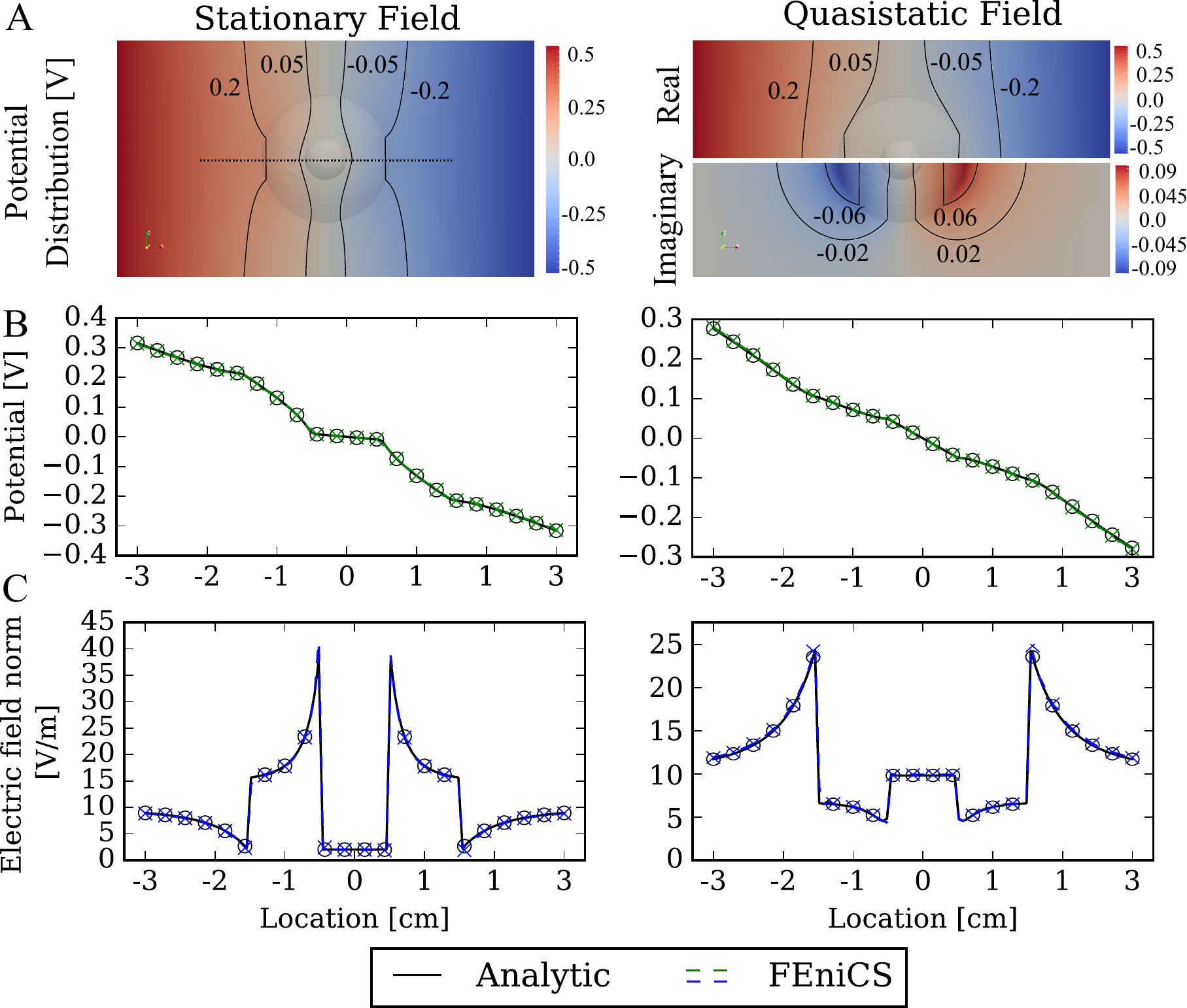,width=1.0\columnwidth} }
    \caption{Comparison of the electric potential and electric field norm obtained by the computational and analytical field solution for the field validation model (Fig. \ref{fig:analyticalproblemgeometry}). A: The potential distribution for the stationary field  and quasistatic equation. The real and imaginary part of the electric potential for the quasistatic equation are shown in the top and bottom of the image, respectively. Potential isolines between \SI{-200}{mV} and \SI{200}{mV} are shown. B, C: Comparison of the electric potential and electric field norm along a cut line through the layered sphere for the computational and the analytical solution for the stationary field and quasistatic field equation. For the quasistatic field equation, the norm of the electric potential multiplied by the sign of its real part is shown.}
    \label{fig:resultanalyticalmodel}
\end{figure} 
\begin{figure}[t]
    \centerline{\psfig{figure=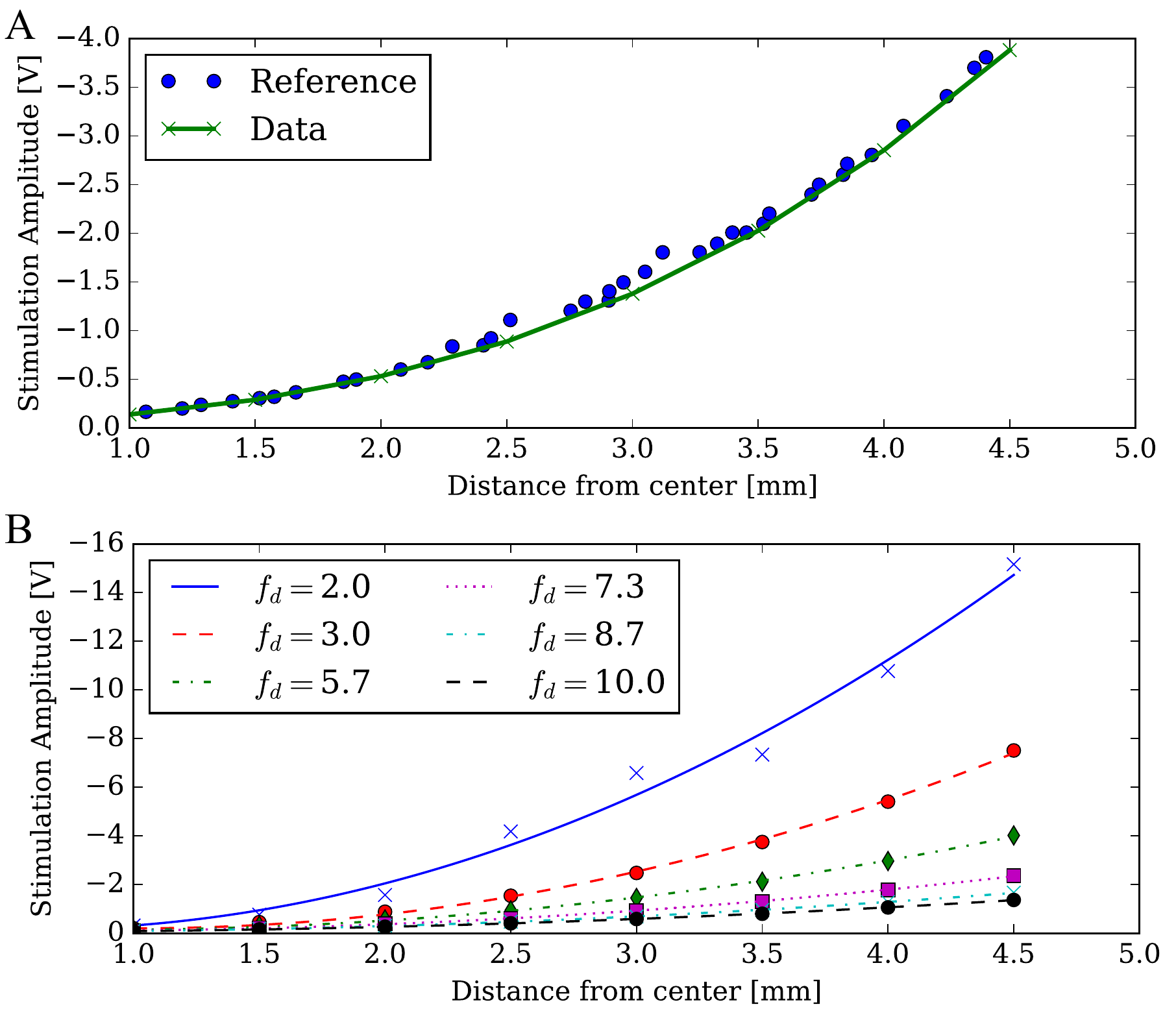,width=1.0\columnwidth} }
    \caption{A: Comparison of the voltage-distance relationship computed for a \SI{5.7}{\micro m} axon with the reference data provided by \cite[Fig. 2]{mcintyre2004}. \revre{B: Threshold-distance relationships illustrated by quadratic polynomials fitted to the data for axon models with varying fiber diameter $f_d$ in $\mu m$.}}
    \label{fig:resultmcintyrevalidation}
\end{figure}
In order to validate the field solution obtained by the implemented pipeline, a simplified multi-compartment volume conductor model for which the analytical solution is known is used. The analytical example model represents the problem of determining the electric potential distribution in a homogeneous electric field, which is locally disturbed by a conducting layered sphere (Fig. \ref{fig:analyticalproblemgeometry}). Following the modeling and simulation pipeline described in the methods section, a 3D volume conductor model was composed with a radius for the inner sphere of \SI{0.5}{cm} and \SI{1.5}{cm} for the outer sphere. The bounding box around the layered sphere is determined by an edge length of \SI{10}{cm}. A homogeneous electric field of \SI{10}{Vm^{-1}} was generated by applying Dirichlet boundary conditions of \SI{0.5}{V} and \SI{-0.5}{V} on the opposing faces of the bounding box in the $yz-$plane. The conductivity was set to \SI{2.0}{Sm^{-1}} for the inner sphere, \SI{0.1}{Sm^{-1}} for the outer sphere, and \SI{1.0}{Sm^{-1}} for the bounding box. For the quasistatic field problem, relative permittivities of $120$ for the inner sphere, $2\cdot 10^6$ for the outer sphere, and $80$ for the bounding box, for a frequency of \SI{35}{kHz} were applied.  The mesh was manually refined at the spheres and bounding box surfaces, resulting in a total number of approximately $1.1$ million cells. The analytic solution is determined by solving the field equations (\ref{eq:laplace}) analytically using a separation approach, which is explained in detail in the appendix \ref{sec:appendix_analytic}. \rev{The relative deviation of the solution, determined as the norm of the respective field quantity between the computational model and the analytical model along a cut line through the layered sphere was below $7.0\cdot10^{-3}$ and $2.9\cdot10^{-2}$ for the electric potential and electric field norm of the stationary field equation and below $1.1\cdot10^{-2}$ and $2.1\cdot10^{-2}$ for the norm of the complex-valued electric potential and the electric field norm of the quasistatic field equation (Fig. \ref{fig:resultanalyticalmodel}).}

\subsection{Validation of the neural activation solution}
\begin{figure*}[t]
    \centerline{\psfig{figure=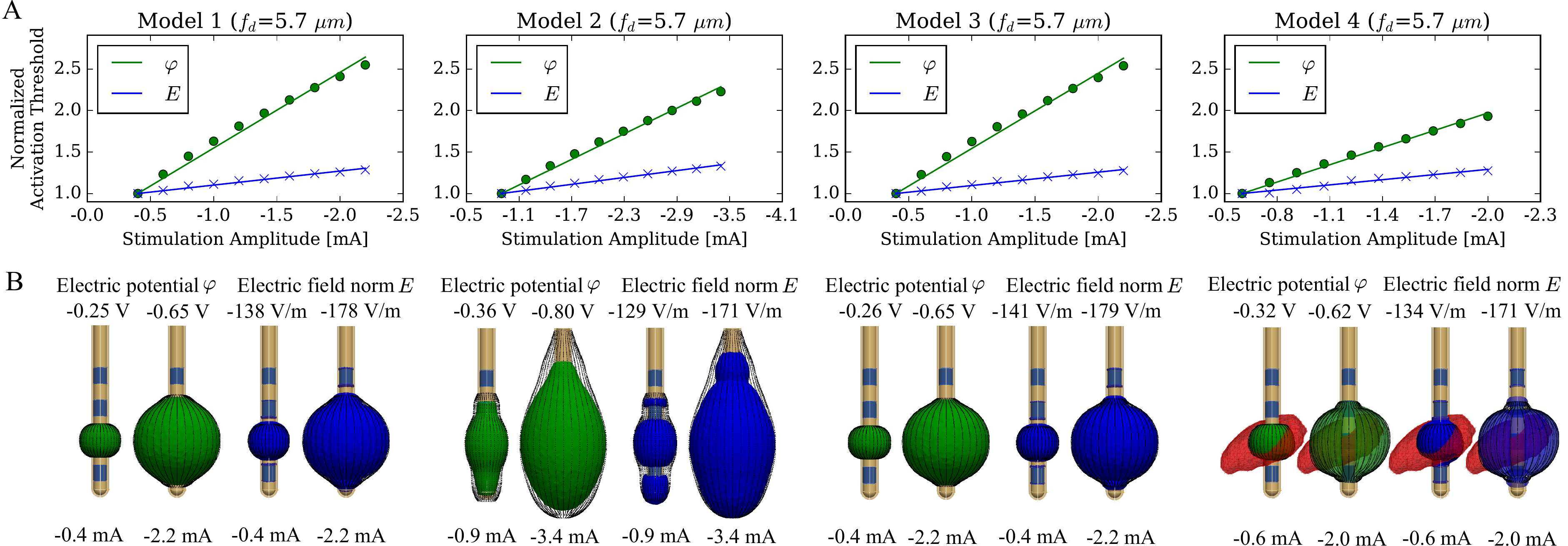,width=1.0\textwidth} }
    \caption{Approximation of the neural activation extent by threshold values of the electric potential $\varphi$ and the electric field norm $E$. A: Normalized activation thresholds for the different models. B: Neural activation extent (shown in black) and iso-volumes of the activation threshold for the electric potential (green) and electric field norm (blue) for the minimum and maximum stimulation amplitude required to activate neural tissue in a distance between \SI{2}{mm} and \SI{4}{mm} \rev{radial} to the active electrode contact center. For Model 4, the STN is illustrated in red.}
    \label{fig:approx_57_with_vtas}
\end{figure*}
To assess the validity of the implementation of the simulation pipeline to determine the neural activation with the myelinated axon cable model \cite{mcintyre2002}, the volume conductor model for DBS used in \cite{mcintyre2004a} is adapted in order to compare the distance-threshold relation for a fiber diameter of \SI{5.7}{\micro m}. The volume conductor model used in the mentioned study comprises the same DBS electrode as in this study as well as an encapsulation layer. A voltage-controlled stimulation signal with a frequency of \SI{150}{Hz} and a pulse duration of \SI{100}{\micro s} is applied. To match the dielectric tissue properties, the conductivity of brain tissue and of the STN was set to \SI{0.3}{Sm^{-1}} and the conductivity of the encapsulation layer to \SI{0.15}{Sm^{-1}}. \rev{The relative deviation between the threshold-distance \rev{relationship} was determined to a value below \SI{5.5}{\percent} by computing the norm between the thresholds determined with the implemented simulation pipeline and the data extracted from \cite[Fig. 2]{mcintyre2004a}, which presents a good agreement of the computed threshold-distance relationship with the data from the original model.} Varying the fiber diameter in the considered range described in section \ref{sec:approxvta} resulted in the expected characteristic decrease in the threshold-distance \rev{relationships} with increasing fiber diameter (Fig. \ref{fig:resultmcintyrevalidation}).

\subsection{Approximating the neural activation by field thresholds}
\begin{figure}[t]
    \centerline{\psfig{figure=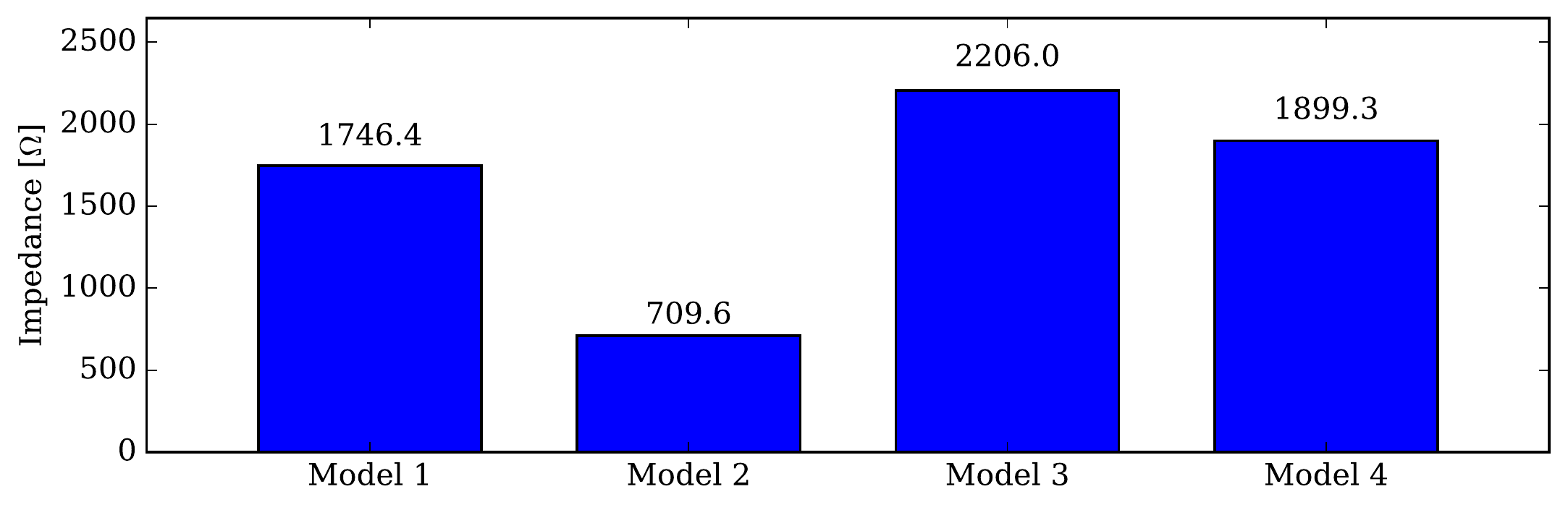,width=1.0\columnwidth} }
    \caption{Electrode impedance determined for the homogeneous model (Model 1), high encapsulation conductivity model (Model 2), low encapsulation conductivity model (Model 3), and the model including the STN as target area (Model 4).}
    \label{fig:impedance}
\end{figure}
The adaptive scheme proposed in \ref{sec:approxvta} was used to determine the neural activation extent for four model setups with varying electrical properties: A homogeneous model (Model 1), a high encapsulation conductivity model (Model 2), a low encapsulation conductivity model (Model 3), and a model including the STN as target area (Model 4). The resulting stimulation amplitude ranges, required to activate a region between \SI{2}{mm} and \SI{4}{mm} distance to the active electrode contact, and the corresponding neural activation extents showed a variation for the different models, which results from their varying electrical tissue properties (Fig. \ref{fig:approx_57_with_vtas}). This influence is also noticeable in the determined activation thresholds for the electric potential with a threshold value of \SI{-0.25}{V} for the minimum stimulation extent with a normalized increase of $2.64$ for the homogeneous model and a threshold value of \SI{-0.32}{V} with a normalized increase of $1.97$ for the model including the STN. The electric field norm threshold value for the minimum stimulation extent varied between \SI{-129}{Vm^{-1}} and \SI{-141}{Vm^{-1}}, but showed a similar normalized increase between $1.29$ and $1.35$ for the models. The varying electrical tissue properties of the models and the resulting varying neural activation extents correspond to changes in the electrode impedance with the high encapsulation conductivity model (Model 2) showing a substantially smaller impedance and also a substantially larger neural activation extent than the other models (Fig. \ref{fig:impedance}).\newline \indent
\begin{figure}[t]
    \centerline{\psfig{figure=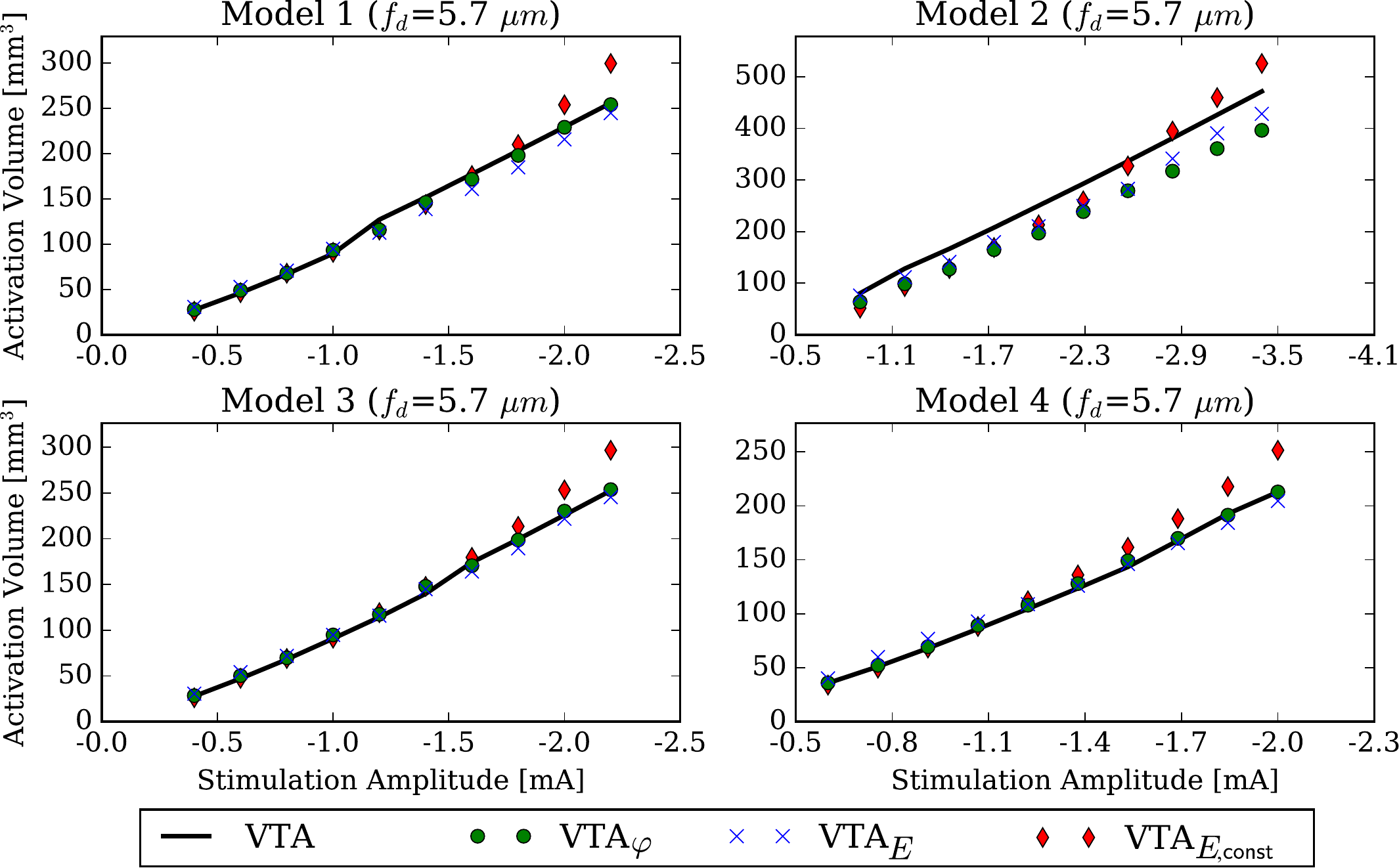,width=1.0\columnwidth} }
    \caption{Volume of the neural activation extent (VTA) and the iso-volume extent for corresponding threshold values of the electric potential (VTA$_\varphi$ in green) and the electric field norm (VTA$_E$ in blue) as well as the approach using the constant threshold value of the electric field norm (VTA$_{E,\mathsf{const}}$ in red) for the different models.}
    \label{fig:fieldapprox}
\end{figure}
\begin{table}[!t]
	\centering
  \renewcommand{\arraystretch}{1.3}
	\caption{\label{tab:fieldvta}Stimulation amplitudes and normalized activation thresholds for approximating the neural activation extent in a distance of \SI{2}{mm} to \SI{4}{mm} \rev{radial} to the active electrode contact.}
	\begin{tabular}{lccccc}
			\hline
			\bfseries Study Case & \bfseries Axon & \bfseries Amplitude & \multicolumn{2}{c}{\bfseries Maximum} \\
			\bfseries  & \bfseries diameter & \bfseries range & \multicolumn{2}{c}{\bfseries normalized threshold} \\
			\bfseries & \bfseries in \SI{}{\micro m} & \bfseries in -\SI{}{mA} & \bfseries for $\varphi$ & \bfseries for $||\boldsymbol E||_2$ \\
			Model 1 		& \rev{2.0} & \rev{[1.2, 8.3]} & \rev{3.37} & \rev{1.70} \\
									& \rev{3.0} & \rev{[0.7, 4.1]} & \rev{2.91} & \rev{1.49} \\
									& 5.7 & [0.4, 2.2] & 2.64 & 1.30 \\
									& 7.3 & [0.3, 1.3] & 2.18 & 1.11\\
								  & 8.7 & [0.2, 0.9] & 2.13 & 1.02\\
									& 10.0 & [0.2, 0.8] & 1.96 & 0.96\\
									\\
			Model 2 		& \rev{2.0} & \rev{[2.6, 13.0]} & \rev{2.98} & \rev{1.74} \\
									& \rev{3.0} & \rev{[1.4, 6.3]} & \rev{2.69} & \rev{1.57}\\
									& 5.7 & [0.9, 3.4] & 2.29 & 1.35 \\
									& 7.3 & [0.6, 2.0] & 2.02 & 1.19\\
								  & 8.7 & [0,4, 1.4] & 2.01 & 1.11\\
									& 10.0 & [0.4, 1.1] & 1.72 & 1.04\\
									\\
			Model 3 		& \rev{2.0} & \rev{[1.2, 8.1]} & \rev{3.30} & \rev{1.67} \\
									& \rev{3.0} & \rev{[0.6, 4.0]} & \rev{3.14} & \rev{1.54}\\
									& 5.7 &	[0.4, 2.2] & 2.63 & 1.29 \\
									& 7.3 & [0.3, 1.3] & 2.18 & 1.11 \\
								  & 8.7 & [0.2, 0.9] & 2.12 & 1.01 \\
									& 10.0 & [0.2, 0.8] & 1.95 & 0.96 \\
									\\
			Model 4 		& \rev{2.0} & \rev{[1.8, 6.0]} & \rev{2.11} & \rev{1.47} \\
									& \rev{3.0} & \rev{[0.9, 3.3]} & \rev{2.18} & \rev{1.67}\\
									& 5.7 & [0.6, 2.0] & 1.97 & 1.29 \\
									& 7.3 & [0.4, 1.2] & 1.79 & 1.18\\
								  & 8.7 & [0.3, 0.9] & 1.73 & 1.11\\
									& 10.0 & [0.2, 0.7] & 1.83 & 1.00\\
			\hline
	\end{tabular}	
\end{table}
In order to assess the quality of approximating the neural activation extent by threshold values of the electric potential and the electric field norm, the volumes of the neural activation extent and the extents resulting from the determined threshold values were compared (Fig. \ref{fig:fieldapprox}). While the extents for the neural activation and for the threshold values were in good agreement for the homogeneous model (Model 1), the low encapsulation conductivity model (Model 3), and the model including the STN as target area (Model 4) \rev{with average volume deviations of \mbox{\SI{4.0}{\percent}$\,\pm\,$\SI{1.8}{\percent}}}, the extents for the threshold values understimated the neural activation extent for the high encapsulation conductivity model (Model 2) \rev{with average volume deviations of \mbox{\SI{14.1}{\percent}$\,\pm\,$\SI{2.8}{\percent}}}, as also noticeable in Figure \ref{fig:approx_57_with_vtas}B for the minimum and maximum stimulation amplitude. \rev{Besides the approach using the threshold-distance relationship to determine the field threshold values, the approach using the mean value of the determined normalized field threshold values as carried out in \cite{astrom2015} was used in this study to approximate the neural activation extent by threshold values of the electric field norm. Comparing both approaches for all models by computing the relative deviation between the approximated activation volumes and the neural activation extents, the approach using the threshold-distance relationship ($\mathsf{VTA}_\mathsf{E}$) showed smaller average volume deviations of \mbox{$8.3\pm3.9$\SI{}{\percent}} (Min: \SI{3.2}{\percent}, Max: \SI{15.6}{\percent}) compared to \mbox{$8.8\pm6.5$\SI{}{\percent}} (Min: \SI{1.0}{\percent}, Max: \SI{22.2}{\percent}) for the approach using the average value of the normalized field thresholds ($\mathsf{VTA}_\mathsf{E,const}$).}\newline \indent
Changing the fiber diameter from \SI{5.7}{\micro m} \rev{to smaller and larger diameters resulted in a variation of the determined stimulation amplitude ranges. The \revre{required} amplitudes to activate the prescribed distance of \SI{2}{mm} to \SI{4}{mm} decreased for increasing fiber diameter, which is also noticeable in the threshold-distance relationship in Figure \ref{fig:resultmcintyrevalidation}B. The largest increases in the normalized activation thresholds were found for small fiber diameters with values up to $3.37$ for the electric potential and $1.74$ for the electric field norm.} With increasing fiber diameter a smaller increase of the normalized activation thresholds for the electric potential and the electric field norm is noticeable, which corresponds to the results reported in \cite{astrom2015}. The \rev{slope} for the normalized activation thresholds for the electric field norm tends to a value of $1$, which allows for approximation of the neural activation extent by using only one threshold value of the electric field norm at an arbitrary stimulation amplitude.\newline \indent
The proposed scheme for the adaptive estimation of the neural activation extent during DBS required approximately \SI{31}{\percent} to \SI{43}{\percent} less axons than a non-adaptive approach, where a prescribed number of axons is positioned in a rectangular grid in each rotational plane (Table \ref{tab:speedup}). While the estimation whether an axon is activated or not activated by the field for a given stimulation amplitude was carried out by one run of the neuron model, the estimation of the minimum stimulation amplitude for one axon required approximately $23.2$ runs for a tolerance of $1\cdot10^{-6}$. Considering the number of required optimization runs, a speed up between \SI{38}{\percent} and \SI{66}{\percent} is obtained by the adaptive scheme, which reduced the computation time for the rotationally symmetric neural activation computations to $24$ - $45$ minutes saving between $10$ - $27$ minutes on a $12\times$\SI{2.4}{GHz}, \SI{48}{GB} workstation. The longest computation time was required for the high encapsulation conductivity model (Model 2), since it employed a wide spread of the neural activation extent along the electrode lead due to the highly conductive encapsulation layer. For the heterogeneous case, the adaptive algorithm reduced the computation time from $19$ hours $41$ minutes to $11$ hours $36$ minutes. When the neural activation extent is estimated by the threshold values of the electric potential and the electric field norm, the number of required axons is substantially reduced from determining the activation thresholds at several locations in each plane around the electrode to determining the threshold-distance relationship along a line \rev{radial} to the active electrode contact. This threshold-distance relationship computation required for a maximum distance of \SI{4}{mm} \rev{radial} to the active electrode contact only the computation of the activation threshold of eight axons, requiring approximately 186 runs of the neuron model and a computation time of below four minutes, which is less than the computation time for the field solution, which required approximately five minutes for all models. In case of the heterogeneous model (Model 4), where no rotational symmetry could be exploited, $288$ axons and a computation time of approximately one hour seven minutes was required.

\section{Discussion}
The present paper proposes an adaptive scheme to estimate the neural activation extent during DBS, which is embedded into a Python package using the open-source solutions FEniCS and NEURON. The field solution of the volume conductor model computed with FEniCS as well as the threshold-distance relationship of the axon model computed with NEURON were compared with analytical solutions as well as reference data from literature and show a good agreement \rev{with deviations below \SI{2.9}{\percent} for the field solution and \SI{5.5}{\percent} for the threshold-distance relationship, respectively} (Fig. \ref{fig:resultanalyticalmodel} and Fig. \ref{fig:resultmcintyrevalidation}). The field model is able to compute stationary current fields (purely resistive material properties) as well as electro-quasistatic fields (complex material properties including conductivity and relative permittivity) for heterogeneous and rotationally asymmetric tissue distributions. The support for incorporating complex material properties further allows for computing the time-dependent field solution \rev{dependent on} the dispersive electrical properties of biological tissue for any applied voltage- or current-controlled DBS signal using the Fourier Finite element method \cite{schmidt2016}. The implementation of the field and neuron parts in one Python package made it possible to adaptively estimate the neural activation extent based on the computed field solution and stimulation signal. Instead of solving the field problem and exporting the time-dependent electric potential at the nodes of several axons located at prescribed positions around the electrode lead to determine the minimum stimulation amplitude to elicit an action potential for each axon \cite{grant2010, schmidt2013}, the adaptive scheme positioned axons only in those regions, where a neural stimulation by the given field solution and stimulation amplitude range would occur. With that, the adaptive scheme requires no pre-knowledge on the neural activation extent for the given volume conductor model and model parameters and, in addition, requires substantially less computational resources and time.\newline \indent
The adaptive scheme was applied to \rev{estimate} the neural activation extent for model cases with varying tissue properties and axon diameters. The tissue properties were chosen to represent different post-operative stages during DBS as well as a homogeneous and rotationally asymmetric case, where the STN was explicitly modeled as target area. Compared to a non-adaptive approach under the assumption that the \rev{minimally} required grid size for the axon positions is already known, a speed up of up to \SI{66}{\percent} was achieved by applying the adaptive scheme.\newline \indent
\begin{table*}[!t]
	\centering
  \renewcommand{\arraystretch}{1.3}
	\caption{\label{tab:speedup}Number of required axon models, speed up and computation time for the different study cases (models). The number of axons and the computation times are determined for the adaptive scheme, the non-adaptive scheme, and for approximating the neuronal activation extent (VTA) by field threshold values for axons with \SI{5.7}{\micro m} fiber diameter. The computation time is measured on a $12\times$\SI{2.4}{GHz}, \SI{48}{GB} workstation. $^\dagger$ rotational symmetry was used for computing the neural activation extent. $^*$ Values determined by mapping number of axons and computation time to rectangular grids of same extent.}
	\begin{tabular}{l|ccc|c|cccc}
			\hline
			\bfseries Study Case 	& \multicolumn{3}{c|}{\bfseries Number of Axons} & \bfseries Speed up
														& \multicolumn{4}{c}{\bfseries Computation time} \\
			\bfseries	& \bfseries VTA								& \bfseries VTA 			& \bfseries VTA 						& \bfseries VTA 			& \bfseries field 			& \bfseries VTA 								& \bfseries VTA 				&	\bfseries VTA \\
			\bfseries	& \bfseries non-adaptive$^*$	& \bfseries adaptive 	& \bfseries field threshold & \bfseries adaptive 	& \bfseries solution 		& \bfseries non-adaptive$^*$ 		& \bfseries adaptive 		& \bfseries field threshold \\
			Model 1$^\dagger$ & $152$ 							& $94$ 								& $8$ 											& \SI{54}{\percent}		& \SI{4.9}{\minute}			& \SI{36.7}{\minute}						& \SI{23.8}{\minute}		& \SI{3.5}{\minute}\\
			Model 2$^\dagger$ & $288$ 							& $173$ 							& $8$ 											& \SI{59}{\percent}		& \SI{4.7}{\minute} 		& \SI{71.3}{\minute}						& \SI{44.8}{\minute}		& \SI{3.4}{\minute} \\
			Model 3$^\dagger$ & $136$ 							& $94$ 								& $8$ 											& \SI{38}{\percent}		& \SI{4.8}{\minute} 		& \SI{35.8}{\minute}						& \SI{25.9}{\minute}		& \SI{3.4}{\minute}\\
			Model 4 	& $4{,}896$ 									& $2{,}813$  					& $288$											& \SI{66}{\percent}		& \SI{5.0}{\minute} 		& $1{,}155.8\,$\SI{}{\minute} 	& \SI{696.3}{\minute}		& \SI{67.0}{\minute} \\
			\hline
	\end{tabular}	
\end{table*}
Nevertheless, even with the adaptive scheme the total computation time for determining the neural activation extent can still be substantially larger than for determining the field solution. To investigate possibilities to further reduce the computational expense, a field threshold approach using the relationship between the field solution and the neural activation suggested in \cite{astrom2015} was applied by computing the neural activation along a line \rev{radial} to the center of the active electrode contact and using the resulting threshold-distance \rev{relationship} to determine threshold values of the electric potential and electric field norm. In \cite{astrom2015}, the iso-volumes for these threshold values constituted a good estimate of the neural activation in homogeneous and rotationally symmetric volume conductor models for DBS. \rev{In \cite{astrom2015}, a single-cable axon model was used compared to a double-cable axon model from \cite{mcintyre2002} used in this study. They show a different threshold-distance relationship for the same fiber diameters, with a \SI{3.0}{\micro m} single-cable axon model correlating with a \SI{5.7}{\micro m} double-cable axon model \cite[Fig. 6]{astrom2015}. The different axon models and different applied stimulation amplitude ranges impede a direct comparison of the normalized field threshold values in this study with the data from \cite{astrom2015}. Nevertheless, larger normalized field thresholds were observered in both studies, when using the electric potential compared to the electric field norm for the field threshold computation (Tab. \ref{tab:fieldvta} and \cite[Table I]{astrom2015}). In addition, in both studies, the maximum normalized field thresholds decreased for increasing fiber diameters tending to a value close to $1$ when using the electric field norm ($1.06$ for a \SI{7.5}{\micro m} axon in \cite{astrom2015} and $0.96$ for a \SI{10.0}{\micro m} for Model 1, see Table \ref{tab:fieldvta}).} Furthermore, the results suggest that also for heterogeneous and rotationally asymmetric field distributions with substantially varying electrode impedances (Fig. \ref{fig:impedance}), threshold values and corresponding iso-volumes of the electric potential and the electric field norm generally constitute a good approximation of the neural activation extent (Fig. \ref{fig:fieldapprox}). However, the results revealed as well that in case of a highly conductive encapsulation layer, as in the phase directly after the surgery, the deviation between the extent approximated by field thresholds and the neural activation extent becomes larger (Fig. \ref{fig:approx_57_with_vtas} and Fig. \ref{fig:fieldapprox}). This increased extent is a result of the increased conductivity of the encapsulation layer, which is spatially connected to the active electrode contact, leading to a higher electric field strength and, with that, an increased activity along the electrode. For this case, the neural activation outside the target area is underestimated, which could have possible implications for determining the impact of unwanted side effects. Nevertheless, the activation in the target area is still approximated with a good quality by the determined field thresholds (Fig. \ref{fig:fieldapprox}). \rev{The underestimation of the neural activation for larger distances away from the active electrode by the field threshold approach \revre{may be due in part to} an approximation artifact. While the response of the axon to a stimulus depends on the electric potential distribution along the whole axon, especially on the second derivative of it \cite{rattay1986}, the field threshold approximation approach determines the field value for a given stimulation amplitude only at the center node of the axon. For a cathodic stimulation pulse and an axon positioned centered to the active stimulation electrode in an isotropic homogeneous medium, the maximum depolarization occurs at its center node, which means that the determined field threshold is connected to the value of the second derivative of the electric potential along the axon. For larger distances away from the electrode, the electric potential and its second derivative along the axon attenuates. Since the field thresholds were determined for distances between \SI{2}{mm} and \SI{4}{mm} from the electrode center, this might present an explanation for the strong correlation between iso-volumes of field threshold values and neural activation extents in the target area, but not for larger distances away from the electrode.\newline \indent}
Regarding the given model parameters, the required stimulation distance or the prescribed stimulation amplitude range and whether rotational symmetry can be exploited, the field threshold can be determined by computing the solution of a few hundred to thousand runs of the axon model, which is substantially less than using only spatially distributed axon models in the non-adaptive and adaptive-approach, which requires $10^4$ to $10^5$ runs of the axon model. \rev{The number of required axon model runs directly depends on the neural activation extent. For instance, a larger stimulation amplitude results in a larger neural activation extent, requiring more axon models run to determine its outer shape.} \rev{Besides the reduction of the computational time achieved by the proposed adaptive and field threshold approximation approach, the computation time could also be decreased by employing a larger time step and smaller number of stimulation pulses. In this study, a time-step of \SI{5}{\micro s} based on \cite{mcintyre2002} and a number of \SI{10}{} pulses were used in order to achieve the 1-to-1 ratio in the firing of the axon models with the DBS pulse train. In any case, the accuracy and convergence of the results have to be carefully checked, when larger time steps or shorter pulse trains are used.\newline \indent
The proposed approach reduces the computation time for a model with a rotationally asymmetric field distribution from more than 11 hours to about one hour on a common workstation. Therefore, we belief that this approach has the potential to take the computation of the neural activation extent closer to a real-world application in clinical practice, where (computation) time is an important constraint.
} \newline \indent
In \cite{astrom2015}, the neural activation extent is approximated by a constant field threshold for a stimulation amplitude, which is computed from the normalized activation thresholds for the corresponding field quantity, such as the electric potential and the electric field norm, by determining a mean value from the linear fit of the field thresholds for increasing stimulation amplitudes. Field threshold values determined with this approach are used in several computational studies to estimate the neural activation extent during DBS \cite{alonso2016, hemm2016, dembek2017}. The determined normalized activation thresholds determined in this study with growth factors between \SI{29}{\percent} to \SI{35}{\percent} for a \SI{5.7}{\micro m} \rev{and \SI{47}{\percent} to \SI{74}{\percent} for a \SI{2.0}{\micro m}} axon suggest that this approach leads to an over- and underestimation of the neural activation extent with volume deviations of up to \rev{\SI{24}{\percent}} by using a constant electric field norm threshold for all stimulation amplitudes \rev{(Table \ref{tab:fieldvta})}. Similar to the results in \cite{astrom2015}, the deviations decreased for increasing axon diameter (Table \ref{tab:fieldvta}). Considering that generally smaller axon diameters, such as \SI{5.7}{\micro m} and below, are used to estimate the neural activation extent during DBS \cite{butson2005, grant2010, schmidt2013, sotiropoulos2007}, this approach might lead to substantial deviations in the estimated extents when a constant field threshold is used for its approximation. The proposed approach to determine for each model the threshold-distance \rev{relationship} along a line \rev{radial} to the active electrode contact to determine the corresponding field threshold for the given stimulation protocol was able to estimate the neural activation extents with deviations below \SI{7.6}{\percent} using electric field norm threshold values and below \SI{3.2}{\percent} using electric potential threshold values for the corresponding stimulation amplitudes. The model case representing the acute post-operative phase by a highly conductive encapsulation layer constituted the exception to the approximation quality with deviations of \SI{11.9}{\percent} and \SI{17.5}{\percent}, respectively, which can be accounted to the spread of the neural activation extent along the electrode lead (Fig. \ref{fig:approx_57_with_vtas}). Nevertheless, in contrast to the constant field threshold approach, the suggested threshold-distance field threshold approach ensures that the activation distance in the target region and with that the neural activation extent in the target region equals the extent computed with solely using the axon models. Therefore, the results suggest that this approach is feasible to estimate the neural activation extent \rev{dependent on} varying dielectric tissue properties, especially for chronic post-operative phases with a low conductive encapsulation layer, and of varying axon diameters while reducing substantially the computational expense.\newline \indent
The axon models to determine the neural activation extent are distributed along normal trajectories in planes around the stimulation electrode, which is a common positioning used for the estimation of the neural activation \cite{butson2005, grant2010, schmidt2013}. This positioning scheme was used in this study in order to compare the simulation results with data from \cite{astrom2015}. Considering the anatomy of the target nuclei for DBS, additional knowledge on the orientation and topology of the axons in the target area could provide a more target-specific and realisitic estimation of the neural activation. \rev{For instance, the used positioning of the axon models around the electrode is a major simplification compared to the more non-uniform orientation of axonal fibers in and around the STN \cite{axer2011}.} The consideration of mutually varying axon fiber orientations and geometries would require a modification of the scheme to estimate the neural activation extent to determine an activation ratio or percentage in the target area rather than a closed volume.\newline \indent
\rev{The described process of finding the inactivated hull in each plane around the electrode lead accounts for a monopolar electrode configuration, where one axon seed point is placed in front of the active stimulation electrode in each plane. In case of a multipolar electrode setup, axon seed points have to be placed in front of each active stimulation electrode in order to ensure that a closed inactivated hull is determined for the given electrode configuration. The advantage of a multipolar electrode setup is that the stimulation amplitude for each active stimulation electrode can be adjusted to achieve the optimal neural activation extent. The adaptive algorithm to determine the inactivated hull for all possible combinations of stimulation amplitudes of a multipolar electrode setup would have to be adjusted to include the minimum and maximum stimulation amplitude scenario. Even for multipolar electrode configurations, this would allow to profit from the speed up of the proposed adaptive algorithm in comparison to the non-adaptive algorithm. Regarding the approximation of the neural activation extent by field threshold values, this is not generally given: For a monopolar electrode configuration with one active stimulation electrode, the field threshold approach determines from one threshold-distance relationship one distance value and one corresponding field threshold value for a given stimulation amplitude. This is carried out in each plane around the stimulation electrode lead. In case of a multiplolar electrode configuration, the field distribution varies for the given stimulation amplitude at each active stimulation electrode, which results in interaction effects modifying the threshold-distance relationship for each active stimulation electrode. Therefore, for each given stimulation amplitude configuration of a multipolar electrode configuration, the threshold-distance relationships have to be recomputed for each active \revre{stimulation} electrode, which reduces the computational speed up. Nevertheless, such an approach would be still substantially faster than recomputing the whole neural activation extent using axon models by the non-adaptive approach. \revre{Besides these required modifications on the adaptive neural activation algorithm, the implementation of multipolar electrode configuration support requires also the consideration of varying stimulation amplitudes for the active electrode contacts, which will be the focus of future releases of the FanPy Python package.}\newline \indent}
The volume conductor model used in this study is embedded into a Python package, which accounts for stationary as well as electro-quasistatic field problems and, therefore, is able to compute the time-dependent field solution employing for resistive as well as dispersive dielectric properties of biological tissue for voltage-controlled and current-controlled stimulation signals. To date, current-controlled stimulation is the favoured stimulation protocol due to the reduced sensitivity of the stimulation impact regarding the dielectric tissue properties and effects of the electrode-tissue interface compared to voltage-controlled sitmulation \cite{lempka2010}. For the estimation of the time-dependent field solution during voltage-controlled stimulation, the results of previous studies point out the necessity to incorporate the dielectric effects at the electrode-tissue interface, which show a dispersive as well as non-linear behaviour with respect to the intensity of the stimulation signal \cite{richardot2002, cantrell2008, howell2015}. Voltage-controlled stimulation was used in this study to validate the threshold-distance relationship of the implemented axon model \cite{mcintyre2002} neglecting the dielectric effects of the electrode-tissue-interface in order to adapt the model used to generate the reference data \cite{mcintyre2004a}.  The currently embedded volume conductor model \rev{is already} able to account for isotropic heterogeneous and dispersive dielectric tissue properties. Besides tissue heterogeneity, the anisotropic dielectric properties of brain tissue can have a substantial influence on the estimation of the neural activation extent and the prediction of side effects during DBS \cite{howell2016}. Therefore, it is planned to incorporate the support for anisotropic conductivity tensors into the field model for future studies.

\section{Conclusion}
In this study, an adaptive scheme to estimate the neural activation extent during DBS is presented. The computation of the field solution as well as the coupling to axon models and the adaptive computation of their response to the stimulation signal is embedded into an open-source Python package, which was used to estimate the neural activation for varying axon diameters and electrical tissue properties rendering different post-operative stages and target area properties. The determined neural activation extents were used to assess the feasibility of their approximation by field threshold values. By using the threshold-distance relationship for determining the field thresholds and corresponding iso\rev{-}volumes \rev{dependent on} the stimulation amplitude, a \rev{close approximation of} the determined neural activation extents could be achieved, while substantially reducing the computational expense.

\section{APPENDIX}
\label{sec:appendix_analytic}
The field equation (\ref{eq:laplace}) for a conducting layered sphere with radius $R_i$ of the inner sphere and $R_e$ of the outer sphere in an external homogeneous electric field as illustrated in Figure \ref{fig:analyticalproblemgeometry} can be formulated using spherical coordinates and a rotational symmetry with respect to the azimuthal angle $\phi$. Within an homogeneous isotropic medium, the field equation for the given problem has the form
\begin{equation}
\frac{1}{r^2} \frac{\partial}{\partial r} \left( r^2 \frac{\partial \varphi}{\partial r}\right) + \frac{1}{r^2 \sin \theta} \frac{\partial}{\partial \theta} \left(\sin \theta \frac{\partial \varphi}{\partial \theta} \right) = 0
\end{equation}
with the radial distance $r$ and the polar angle $\theta$. Since $r$ and $\theta$ can be varied independently, a separation of the potential $\varphi = R(r)\Theta(\theta)$ results in the angular and radial equations
\begin{eqnarray}
	\frac{1}{\sin \theta} \frac{d}{d \theta} \left(\sin \theta \frac{d \Theta(\theta)}{d \theta} \right) + \lambda \Theta(\theta) &=& 0 \\
	\frac{d}{dr} \left( r^2 \frac{d R(r)}{r}\right) - \lambda R(r) &=& 0
\end{eqnarray}
\rev{with the separation constant $\lambda$,} where the solution to the angular equation is given by 
\begin{equation}
	\Theta_n(\theta) = D_n P_n(\cos \theta), \, P_n(x) = \frac{1}{2^n n!} \frac{d^n}{dx^n}\left((x^2-1)^n\right)
\end{equation}
using the substitution $x = \cos(\theta)$ and \rev{$\lambda = n(n+1)$} with a constant $D_n$ and the Legendre polynomial $P_n(x)$ for $n=0,1,2,\ldots$ \cite{sukhorukov2001}. The radial equation represents an Euler-Cauchy differential equation, which solution is given by
\begin{equation}
	R_n(r) = C_{1,n}r^n + C_{2,n}r^{-(n+1)}
\end{equation}
with the constants $C_{1,n}$ and $C_{2,n}$. Applying the separation equation, the general solution of the potential $\varphi := \varphi(r,\theta)$ is then given by
\begin{equation}
\varphi = \begin{cases} \sum \limits_{n=0}^\infty A_nr^n P_n(x) &,0\leq r \leq R_i \\
												\sum \limits_{n=0}^\infty \left(B_nr^n+C_nr^{-(n+1)}\right)P_n(x) &,R_i<r\leq R_e \\
												-E_0rx + \sum \limits_{n=0}^\infty D_nr^{-(n+1)} P_n(x) &,R_e<r
										\end{cases}
\end{equation}
using the substitution $x = \cos(\theta)$ with constants $A_n, B_n, C_n, D_n$ for an external homogeneous electric field in $z$-direction $-E_0z=-E_0r\cos(\theta)$ and a vanishing influence of the conducting layered sphere for $r\rightarrow\infty$. Applying the continuity conditions
\begin{eqnarray}
	\varphi_1 - \varphi_2 &=& 0 \\
	\boldsymbol n(\boldsymbol J_1 - \boldsymbol J_2) &=& 0
\end{eqnarray}
for the electrical potential $\varphi$ and the current density $\boldsymbol J$ at the boundaries of the layered sphere and exploting the orthnormality of the Legendre polynomials results in a system of linear equations which has only for $n=1$ a non-trivial solution and non-zero right hand side
\begin{equation}
\begin{pmatrix} R_i^3 & -R_i^3 & -1 & 0 \\ \sigma_{i,i} & -\sigma_{s,i} & 2\sigma_s & 0 \\ 0 & R_e^3 & 1 & -1 \\ 0 & \sigma_{s,e} & -2\sigma_s & 2\sigma_e \end{pmatrix} \cdot \begin{pmatrix} A_1 \\ B_1 \\ C_1 \\ D_1 \end{pmatrix} = \begin{pmatrix} 0 \\ 0 \\ -E_{0,e} \\ -\sigma_eE_{0,e} \end{pmatrix}
\label{eq:lse}
\end{equation}
with $\sigma_{i,i} = \sigma_iR_i^3$, $\sigma_{s,i} = \sigma_sR_i^3$, $\sigma_{s,e} = \sigma_sR_e^3$, $E_{0,e} = E_0R_e^3$. The solution of (\ref{eq:lse}) is then given by
\begin{equation}
\varphi(r,\theta) = \begin{cases} A_1r\cos\theta &,0\leq r \leq R_i \\
												\left(B_1r+C_1r^{-2}\right)\cos\theta &,R_i<r\leq R_e \\
												\left(D_1r^{-2}-E_0r\right) \cos \theta &,R_e<r \text{ .}
										\end{cases}
\end{equation}
\bibliographystyle{IEEEtran}
\bibliography{IEEEabrv,references}

\end{document}